\documentclass[10pt]{article}
\usepackage[utf8]{inputenc}
\usepackage{dsfont}
\usepackage{amsfonts}
\usepackage{amsmath}
\usepackage{amssymb}
\usepackage{enumitem}
\usepackage{mathrsfs}
\usepackage{graphicx}
\usepackage[T1]{fontenc}
\usepackage[english]{babel}
\usepackage{fullpage}

\usepackage{xcolor}

\usepackage{tikz-feynman}
\tikzfeynmanset{compat=1.0.0}

\begin{document}

\begin{titlepage}

\begin{center}

\begin{flushright}

\end{flushright}

\begin{flushright}

\end{flushright}

\begin{flushright}

\end{flushright}

{\Large\bf Linear transformations of\\\vspace{0.5cm} Srivastava's $H_C$ triple hypergeometric function\\}

\vspace{1.5cm}

Samuel Friot$^{a,b, \ddagger}$ and Guillaume Suchet-Bernard$^{a}$\\[1.5cm]
\textit{$^a$ Universit\'e Paris-Saclay, CNRS/IN2P3, IJCLab, 91405 Orsay, France } \\[0.5cm]
\textit{$^b$ Univ Lyon, Univ Claude Bernard Lyon 1, CNRS/IN2P3, \\
 IP2I Lyon, UMR 5822, F-69622, Villeurbanne, France}\\[2.cm]
\end{center}

\begin{abstract}

We explore the large set of linear transformations of Srivastava's $H_C$ triple hypergeometric function. This function has been recently linked to the massive one-loop conformal scalar 3-point Feynman integral. We focus here on the class of linear transformations of $H_C$ that can be obtained from linear transformations of the Gauss $_2F_1$ hypergeometric function and, as $H_C$ is also a three variable generalization of the Appell $F_1$ double hypergeometric function, from the particular linear transformation of $F_1$ known as Carlson's identity and some of its generalizations.  These transformations are applied at the level of the 3-fold Mellin-Barnes representation of $H_C$. This allows us to use the powerful conic hull method of Phys. Rev. Lett. \textbf{127} (2021) no.15, 151601 for the evaluation of the transformed Mellin-Barnes integrals, which leads to the desired results. The latter can then be checked numerically against the Feynman parametrization of the conformal 3-point integral.
We also show how  this approach can be used to derive many known (and less known) results involving Appell double hypergeometric functions.

\end{abstract}

\vspace{4cm}

\small{$\ddagger$ samuel.friot@universite-paris-saclay.fr}

\end{titlepage}

\baselineskip 12pt

\section{Introduction}

The transformation theory of multivariable hypergeometric functions can be studied using various methods, as described for instance in Chapter 9 of \cite{Srivastava}. One of the simplest possibilities is to write the studied hypergeometric function of $n$ variables as an infinite sum of hypergeometric series with $n-1$ variables and to apply the known transformation theory of the latter. This approach has been followed in a considerable number of publications (we refer the reader to \cite{Srivastava} for a non-exhaustive list) and it has been revisited recently, in the linear case, by one of the authors of the present paper and collaborators, for the realization of a \textit{Mathematica} package dedicated to the numerical evaluation of the Appell $F_2$ double hypergeometric function \cite{Ananthanarayan:2021bqz}. In addition, a first automatization of this linear transformation procedure has been presented in \cite{Ananthanarayan:2021yar}. Its application to various cases of interest such as other Appell functions, Horn functions, Lauricella functions, etc., with the aim of building \textit{Mathematica} packages for their numerical evaluation, is presently under study  \cite{ABFP, ABF}.

We explore here another approach, based on multifold Mellin-Barnes (MB) integral representations of hypergeometric functions, on which transformations are directly performed, before being calculated analytically. It seems that this approach has been less used in the literature than the other methods described in \cite{Srivastava}. However, an important progress in the theory of multiple MB integrals \cite{Ananthanarayan:2020fhl}, allowing the evaluation of the latter as linear combinations of multivariable series, encouraged us to study its potential ability to derive new results in the transformation theory of multivariable hypergeometric functions. As a first case of study, we choose to focus on Srivastava $H_C$ triple hypergeometric function which, as recently shown in \cite{Loebbert:2020hxk}, is related to the massive one-loop conformal scalar 3-point Feynman integral. Even more interestingly, it was possible to embed $H_C$ in a more general series describing the $N$-point case in \cite{Loebbert:2020glj}. These quantum field theory results, associated with other mathematical expressions of the massive one-loop conformal scalar $N$-point Feynman integral (also obtained in \cite{Loebbert:2020glj} for alternative choices of the conformal variables) resulted in the derivation of a set of new quadratic transformations for the involved multivariable hypergeometric functions, and in particular for $H_C$, in \cite{ABFG2}. A few linear transformations of $H_C$ have also been obtained in \cite{ABFG2}.
In the present work, we go on these investigations and we will be concerned with a large set of other linear transformations of $H_C$, the vast majority of which are new results.

Our method rests on the application of linear transformations of Gauss $_2F_1$ and Appell $F_1$ functions at the level of the 3-fold MB integral representation of $H_C$. The transformed MB integrals can then be computed analytically using the method of \cite{Ananthanarayan:2020fhl} and the resulting linear transformations of $H_C$ numerically checked using the Feynman parametrization of the conformal 3-point integral.

In Section \ref{lin2F1} we give a brief summary of the (non-logarithmic) linear transformations of the Gauss  $_2F_1$ hypergeometric function. In Section \ref{linAppell}, we explain our method and show how to derive many known (and a few less known) linear transformations of the Appell functions. We also present Carlson's identity and related results which will be useful for our analysis, as well as a notation that will ease the classification of the transformed MB integrals studied in this paper. Then, in Section \ref{linHC}, we present our study of the $H_C$ case, the main part of the corresponding results being given, due to their length, in an ancillary \textit{Mathematica} file of the arXiv submission. We finish the paper by a section containing our conclusions.

\section{Linear transformations of the Gauss hypergeometric ${}_2F_1$ function\label{lin2F1}}

The well-known linear transformations of the Gauss hypergeometric ${}_2F_1$ function, listed for instance in \cite{A&S} (see Eqs. 15.3.3 to 15.3.9 for the non-logarithmic cases on which we focus here), can all be obtained from its Mellin-Barnes representation\footnote{In the following, MB contours are such that the set of poles of each Gamma function of the MB integrand is not split in different subsets by the contours.}:
\begin{equation}\label{2F1MB1}
	{}_2F_1(a,b;c;z)=\frac{\Gamma(c)}{\Gamma(a)\Gamma(b)}\int_{-i\infty}^{i\infty} \frac{ds}{2i\pi}(-z)^s\Gamma(-s)
	\frac{\Gamma(a+s)\Gamma(b+s)}{\Gamma(c+s)}
\end{equation}
and from the following linear transformations of the latter integral:
\begin{equation}\label{2F1MB2}
	{}_2F_1(a,b;c;z)=\frac{\Gamma(c)}{\Gamma(a)\Gamma(b)\Gamma(c-a)\Gamma(c-b)}\int_{-i\infty}^{i\infty} \frac{ds}{2i\pi}(1-z)^{s}\Gamma(-s)\Gamma(c-a-b-s)\Gamma(a+s)\Gamma(b+s)
\end{equation}
\begin{equation}\label{2F1MB3}
	{}_2F_1(a,b;c;z)=(1-z)^{c-a-b}\frac{\Gamma(c)}{\Gamma(c-b)\Gamma(c-a)}\int_{-i\infty}^{i\infty} \frac{ds}{2i\pi}(-z)^s\Gamma(-s)\frac{\Gamma(c-a+s)\Gamma(c-b+s)}{\Gamma(c+s)} \end{equation}
\begin{equation}\label{2F1MB4}
	{}_2F_1(a,b;c;z)=(1-z)^{-a}\frac{\Gamma(c)}{\Gamma(a)\Gamma(c-b)}\int_{-i\infty}^{i\infty} \frac{ds}{2i\pi}\left(-\frac{z}{z-1}\right)^s\Gamma(-s)\frac{\Gamma(a+s)\Gamma(c-b+s)}{\Gamma(c+s)} 
\end{equation}
and its symmetrical version
\begin{equation}\label{2F1MB5}
	{}_2F_1(a,b;c;z)=(1-z)^{-b}\frac{\Gamma(c)}{\Gamma(b)\Gamma(c-a)}\int_{-i\infty}^{i\infty} \frac{ds}{2i\pi}\left(-\frac{z}{z-1}\right)^s\Gamma(-s)\frac{\Gamma(b+s)\Gamma(c-a+s)}{\Gamma(c+s)} 
\end{equation}
Indeed, by closing the contour of integration of these Mellin-Barnes integrals to right, one obtains from Eq.(\ref{2F1MB3}) (resp. (\ref{2F1MB4}), (\ref{2F1MB5}) and (\ref{2F1MB2})) Eq.15.3.3 (resp. 15.3.4, 15.3.5 and 15.3.6) of  \cite{A&S}, while closing the contour to the left Eq.(\ref{2F1MB1}) (resp. (\ref{2F1MB2}), (\ref{2F1MB3})) leads to Eq.15.3.7 (resp. 15.3.8 and 15.3.9).

It was explicitly shown a long time ago in \cite{W&W} how to derive Eq.(\ref{2F1MB2}) from Eq.(\ref{2F1MB1}) and, although not mentioned in the latter reference, a similar approach can be used to derive the other MB integrals above from Eq.(\ref{2F1MB2}), using the relations\footnote{Particular values of the variable and/or parameters for which these relations are not true are tacitly omitted.}
\begin{equation}
	(1-z)^{s}\Gamma(c-a-b-s)=(1-z)^{c-a-b}\int_{-i\infty}^{i\infty} \frac{dt}{2i\pi}(-z)^t\Gamma(-t)\Gamma(t+c-a-b-s)	
\end{equation}
or
\begin{equation}
	\left(1-z\right)^{s}\Gamma(a+s)=\left(1-\frac{z}{z-1}\right)^{-s}\Gamma(a+s)=\left(1-z\right)^{-a}\int_{-i\infty}^{i\infty} \frac{dt}{2i\pi}\left(-\frac{z}{z-1}\right)^t\Gamma(-t)\Gamma(t+a+s)	
\end{equation}
followed by an application of the first Barnes lemma in the obtained 2-fold MB integrals.

As we will now show in the next sections, Eqs.(\ref{2F1MB1})-(\ref{2F1MB5}) will be some of the important ingredients in our approach to derive linear transformations of hypergeometric functions with more than one variable, and in particular for the $H_C$ case which is our main object of study in this paper.
es integral representation of the following transformation of the Hypergeometric function :

\section{Linear transformations of Appell double hypergeometric functions from their Mellin-Barnes representations \label{linAppell}}

\subsection{A few examples for the Appell $F_1$ function}
We begin the presentation of our calculational method with the derivation of some linear transformations of the Appell $F_1$ double hypergeometric function, obtained from its Mellin-Barnes representations and transformations of the latter. The $F_1$ function has a particular place in our study as it is a simple two-variable extension of the Gauss $_2F_1$ which, in turn, can be extended at three variables by $H_C$. 

The MB representation of the Appell $F_1$ function is \cite{KdF}
\begin{equation}
	F_1(a,b,b';c;x,y)=\frac{\Gamma(c)}{\Gamma(a)\Gamma(b)\Gamma(b')}\int_{-i\infty}^{i\infty}\frac{dz_1}{2i\pi}
	\int_{-i\infty}^{i\infty}\frac{dz_2}{2i\pi}(-x)^{z_1}(-y)^{z_2}	
	\Gamma(-z_1)\Gamma(-z_2)\frac{\Gamma(a+z_1+z_2)\Gamma(b+z_1)\Gamma(b'+z_2)}{\Gamma(c+z_1+z_2)}\label{F1MB1}
\end{equation}
As recalled in \cite{Ananthanarayan:2020fhl}, it is easy to obtain, directly from this integral, a few of the well-known linear transformations of $F_1$ involving the $G_2$ Horn double hypergeometric function. 

Let us now rewrite the $z_2$ integral as
\begin{equation}
\label{example1}
	\int_{-i\infty}^{i\infty}\frac{dz_2}{2i\pi}(-y)^{z_2}	
	\Gamma(-z_2)\frac{\Gamma(a+z_1+z_2)\Gamma(b'+z_2)}{\Gamma(c+z_1+z_2)}=\frac{\Gamma(a+z_1)\Gamma(b')}{\Gamma(c+z_1)} {_2F_1}(a+z_1,b';c+z_1;y)
\end{equation}
One can now apply any of the Eqs.(\ref{2F1MB2})-(\ref{2F1MB5}) on the r.h.s. of Eq.(\ref{example1}) to obtain a transformed MB integral for $F_1$. For instance, let us choose Eq.(\ref{2F1MB5}). In this case, one obtains
\begin{multline}
		F_1(a,b,b';c;x,y) = \frac{\Gamma(c)}{\Gamma(a)\Gamma(b)\Gamma(b')\Gamma(c-a)}(1-y)^{-b'}\int_{-i\infty}^{i\infty}\frac{dz_1}{2i\pi}\int_{-i\infty}^{i\infty}\frac{dz_2}{2i\pi}
		(-x)^{z_1}\left(-\frac{y}{y-1}\right)^{z_2}\Gamma(-z_1)\Gamma(-z_2) \\
		\times \frac{\Gamma(a+z_1)\Gamma(c-a+z_2)\Gamma(b+z_1)\Gamma(b'+z_2)}{\Gamma(c+z_1+z_2)}\label{F1MB2}
\end{multline}
One can recognize here in the r.h.s. the MB representation of the Appell $F_3$ function \cite{KdF}, which thus gives the well-known transformation (see Eq.(29) of \cite{KdF}, or Eq.(104) of \cite{Srivastava})
\begin{equation}
	F_1(a,b,b';c;x,y)=(1-y)^{-b'}F_3\left(a,c-a,b,b';c;x,\frac{y}{y-1}\right)\label{F1F3}
\end{equation}
As a second example, one can go on the transformation process, choosing to act on the $z_1$ integral in Eq.(\ref{F1MB2}) using still Eq.(\ref{2F1MB5}):
\begin{multline}
\label{example2}
	\int_{-i\infty}^{i\infty}\frac{dz_1}{2i\pi}(-x)^{z_1}	
	\Gamma(-z_1)\frac{\Gamma(a+z_1)\Gamma(b+z_1)}{\Gamma(c+z_1+z_2)}=\frac{\Gamma(a)\Gamma(b)}{\Gamma(c+z_2)} {_2F_1}(a,b;c+z_2;x)\\
	=(1-x)^{-b}\frac{\Gamma(a)\Gamma(c+z_2)}{\Gamma(c-a+z_2)}\int_{-i\infty}^{i\infty}\frac{dz_1}{2i\pi}\left(-\frac{x}{x-1}\right)^{z_1}	
	\Gamma(-z_1)\frac{\Gamma(c-a+z_1+z_2)\Gamma(b+z_1)}{\Gamma(c+z_1+z_2)}
\end{multline}
One then obtains
\begin{multline}
	F_1(a,b,b';c;x,y)=\frac{\Gamma(c)}{\Gamma(b)\Gamma(b')\Gamma(c-a)}(1-x)^{-b}(1-y)^{-b'}\int_{-i\infty}^{i\infty}\frac{dz_1}{2i\pi}
	\int_{-i\infty}^{i\infty}\frac{dz_2}{2i\pi}\left(-\frac{x}{x-1}\right)^{z_1}\left(-\frac{y}{y-1}\right)^{z_2}\\
	\times\Gamma(-z_1)\Gamma(-z_2)\frac{\Gamma(c-a+z_1+z_2)\Gamma(b+z_1)\Gamma(b'+z_2)}{\Gamma(c+z_1+z_2)}\label{F1MB3}
\end{multline}
which is nothing but the MB representation of one of the five well-known Euler transformations of $F_1$ (see Eq.($5_1$) in \cite{KdF}, or Eq.(1) p.240 in \cite{Erdelyi})
\begin{equation}
	F_1(a,b,b';c;x,y)=(1-x)^b(1-y)^{-b'}F_1\left(c-a,b,b';c;\frac{x}{x-1},\frac{y}{y-1}\right)
\end{equation}
Trying all possible ways to transform the MB representation of $F_1$ using Eqs.(\ref{2F1MB2})-(\ref{2F1MB4}), one can derive many known (and less known) results, which will be fully listed elsewhere \cite{ABF}, such as those coming from the following transformed MB representation :
\begin{multline}
	F_1(a,b,b';c;x,y) = \frac{(1-x)^{-a}\Gamma(c)}{\Gamma(a)\Gamma(b')\Gamma(c-a)\Gamma(c-b'-b)}\int_{-i\infty}^{i\infty}\frac{dz_1}{2i\pi}\int_{-i\infty}^{i\infty} \frac{dz_2}{2i\pi}\left(\frac{-x}{x-1}\right)^{z_1}\left(\frac{1-y}{1-x}\right)^{z_2}\Gamma(-z_1)\Gamma(-z_2) \\
	\times
	\frac{\Gamma(a+z_1+z_2)\Gamma(c-b'-b+z_1)\Gamma(b'+z_2)\Gamma(c-a-b'-z_2)}{\Gamma(c-b'+z_1)}\label{F1MB4}
\end{multline}
Evaluating this integral with the method of \cite{Ananthanarayan:2020fhl} provides several linear transformations of $F_1$ among which 
\begin{multline}
	F_1(a,b,b';c;x,y) = (1-x)^{-a}\frac{\Gamma(c)\Gamma(c-a-b')}{\Gamma(c-a)\Gamma(c-b')}F_2\left(a,c-b-b',b';c-b',1-c+a+b';\frac{x}{x-1},\frac{y-1}{x-1} \right)\\
	+(1-x)^{b'-c}(1-y)^{c-a-b'}\frac{\Gamma(c)\Gamma(a+b'-c)}{\Gamma(b')\Gamma(a)}F_2\left(c-b',c-b-b',c-a;c-b',1+c-a-b';\frac{x}{x-1},\frac{y-1}{x-1} \right)
\end{multline}
involving two Appell $F_2$ functions and that we could not find in any of the quoted references. The same integral also produces the following results \begin{multline}
	F_1(a,b,b';c;x,y) = (1-x)^{b'-a}(1-y)^{-b'}\frac{\Gamma(c)\Gamma(a-b')}{\Gamma(a)\Gamma(c-b')}H_2\left(a-b',c-b-b',b',c-a;c-b';\frac{x}{x-1},\frac{y-1}{1-x} \right) \\
	+(1-y)^{-a}\frac{\Gamma(c)\Gamma(b'-a)}{\Gamma(b')\Gamma(c-a)}F{}^{2:1;0}_{1:1;0}
  \left[
   \setlength{\arraycolsep}{0pt}
   \begin{array}{c@{{}:{}}c@{;{}}c}
  a, c-b' & c-b-b' & - \\[1ex]
   1-b'+a & c-b' & -
   \end{array}
   \;\middle|\;
 \frac{x}{1-y},\frac{x-1}{y-1}
 \right]
 \end{multline}
where $F{}^{2:1;0}_{1:1;0}$ is a Kamp\'e de F\'eriet function \cite{KdF,Srivastava}, and

\begin{multline}
	F_1(a,b,b';c;x,y) = (1-x)^{-a}\left(\frac{-x}{x-1}\right)^{b+b'-c} \\
	\times \frac{\Gamma(c)\Gamma(c-a-b')\Gamma(a+b+b'-c)}{\Gamma(a)\Gamma(c-a)\Gamma(b)}H_2\left(a+b+b'-c,b',1-b,c-b-b';1-c+a+b';\frac{y-1}{x-1},\frac{1-x}{x} \right) \\
	+ (1-y)^{-a}\left(\frac{-x}{x-1}\right)^{b}\left(\frac{-x}{y-1}\right)^{b'-c}\frac{\Gamma(c)\Gamma(a+b'-c)}{\Gamma(a)\Gamma(b')}H_2\left(b,c-a,1-b,c-b-b';1+c-a-b';\frac{y-1}{x-1},\frac{1-x}{x} \right) \\
	+ x^{-a}\frac{\Gamma(c)\Gamma(c-a-b-b')}{\Gamma(c-b-b')\Gamma(c-a)}F{}^{2:1;0}_{1:1;0}
  \left[
   \setlength{\arraycolsep}{0pt}
   \begin{array}{c@{{}:{}}c@{;{}}c}
  a, 1-c+a+b' & b' & - \\[1ex]
   1-c+a+b+b' & 1-c+a+b' & -
   \end{array}
   \;\middle|\;
 \frac{1-y}{x},\frac{x-1}{x}
 \right]
\end{multline}
The spectrum of accessible transformations of the MB representation of $F_1$ can even be enlarged using adequate changes of variables, as shown in the following examples.

Applying the change of variables $z_1=s_1$ and $z_2=-a-s_1-s_2$ to Eq.(\ref{F1MB1}), followed by the transformation of the MB integral over $s_1$ (which has the form of Eq.(\ref{2F1MB2})) using Eq.(\ref{2F1MB1}), one obtains after inverse change of variables
\begin{multline}
	F_1(a,b,b';c;x,y)=\frac{\Gamma(c)}{\Gamma(a)\Gamma(b)}\int_{-i\infty}^{i\infty}\frac{dz_1}{2i\pi}\int_{-i\infty}^{i\infty}\frac{dz_2}{2i\pi}
	(y-x)^{z_1}(-y)^{z_2}\Gamma(-z_1)\Gamma(-z_2)\\
	\times\frac{\Gamma(a+z_1+z_2)\Gamma(b+b'+z_1+z_2)\Gamma(b+z_1)}{\Gamma(c+z_1+z_2)\Gamma(b+b'+z_1)}\label{CarlsonMB}
\end{multline}
which, up to the overall ratio of Gamma functions, is the MB representation of Carlson's identity \cite{Carlson}
\begin{equation}
	\sum_{m,n=0}^\infty\frac{
	x^{m}y^{n}}{m!n!} 
	\frac{(a)_{m+n}(b)_{m}(b')_n}{(c)_{m+n}}=\sum_{m,n=0}^\infty\frac{
	(x-y)^{m}y^{n}}{m!n!} 
	\frac{(a)_{m+n}(b'+b)_{m+n}(b)_m}{(c)_{m+n}(b'+b)_m}
\end{equation}
One will note that interchanging $z_1$ with $z_2$ in the calculations leading to Carlson's identity gives symmetrical expressions which reflect the symmetry $F_1(a,b,b';c;x,y)=F_1(a,b',b;c;y,x)$.
 
In addition to this identity, the computation of the MB integral of Eq.(\ref{CarlsonMB}) following the method of \cite{Ananthanarayan:2020fhl} gives 
\begin{multline}
	F_1(a,b,b';c;x,y)=(-y)^{-a}\frac{\Gamma(c)\Gamma(b+b'-a)}{\Gamma(c-a)}F_2\left(a;b,1+a-c;b+b',1+a-b-b';\frac{y-x}{y},\frac{1}{y}\right)\\
	+(-y)^{-b-b'}\frac{\Gamma(c)\Gamma(a-b-b')}{\Gamma(a)\Gamma(c-b-b')}F_2\left(b+b':b,1-c+b+b';b+b',1-a+b+b';\frac{y-x}{y},\frac{1}{y}\right)
\end{multline}
\begin{multline}
	F_1(a,b,b';c;x,y)=(y-x)^{-a}\frac{\Gamma(c)\Gamma(b-a)}{\Gamma(b)\Gamma(c-a)}F{}^{2:0;1}_{1:0;1}
  \left[
   \setlength{\arraycolsep}{0pt}
   \begin{array}{c@{{}:{}}c@{;{}}c}
  a, 1+a-b-b' & - & 1+a-c \\[1ex]
   1+a-b & - & 1+a-b-b'
   \end{array}
   \;\middle|\;
 \frac{y}{y-x},\frac{1}{x-y}
 \right]
\\
	+(-x+y)^{-b_1}\frac{\Gamma(c)\Gamma(a-b)}{\Gamma(a)\Gamma(c-b_1)}\sum_{m,n=0}^\infty y^m\left(\frac{1}{y-x}\right)^n\frac{(a-b)_{m-n}(b')_{m-n}(1-b')_n(b)_n}{m!n!(c-b)_{m-n}}
\end{multline}
and
\begin{multline}
	F_1(a,b,b';c;x,y)=(y-x)^{-a}\frac{\Gamma(c)\Gamma(b-a)}{\Gamma(b)\Gamma(c-a)}F{}^{2:0;1}_{1:0;1}
  \left[
   \setlength{\arraycolsep}{0pt}
   \begin{array}{c@{{}:{}}c@{;{}}c}
  a, 1+a-b-b' & - & 1+a-c \\[1ex]
   1+a-b & - & 1+a-b-b'
   \end{array}
   \;\middle|\;
 \frac{y}{y-x},\frac{1}{x-y}
 \right]
\\
	+(-y)^{-a+b}(-x+y)^{-b}\frac{\Gamma(c)\Gamma(-a+b+b')\Gamma(a-b)}{\Gamma(a)\Gamma(b')\Gamma(c-a)}H_2\left(a-b:1+a-c;b,1-b';1+a-b-b';\frac{1}{y},\frac{y}{x-y}\right)\\
	+(-y)^{-b'}(-x+y)^{-b}\frac{\Gamma(c)\Gamma(a-b-b')\Gamma(b')}{\Gamma(a)\Gamma(b')\Gamma(c-a)}H_2\left(b':1-c+b+b';b,1-b';1-a+b+b';\frac{1}{y},\frac{y}{x-y}\right)
\end{multline}
It is possible to apply the same changes of variables to the more general MB representation  which will be used later in this paper:
\begin{equation}
	I_{p,q}=\int_{-i\infty}^{i\infty}\frac{dz_1}{2i\pi}
	\int_{-i\infty}^{i\infty}\frac{dz_2}{2i\pi}(-x)^{z_1}(-y)^{z_2}	
	\Gamma(-z_1)\Gamma(-z_2)\frac{\Pi_{i=1}^p\Gamma(a_i+z_1+z_2)}{\Pi_{j=1}^q\Gamma(c_j+z_1+z_2)}\Gamma(b+z_1)\Gamma(b'+z_2)\label{F1typeMB}
\end{equation}
When $p$ and $q$ are such that the MB integral is well-defined, this integral generalizes the Appell $F_1$ MB representation (and is thus called "$F_1$-type" in the following).

The corresponding transformed MB integral, called "$KdF$-type", is
\begin{equation}
	I_{p,q}=\Gamma(b')\int_{-i\infty}^{i\infty}\frac{dz_1}{2i\pi}
	\int_{-i\infty}^{i\infty}\frac{dz_2}{2i\pi}(-x+y)^{z_1}(-y)^{z_2}	
	\Gamma(-z_1)\Gamma(-z_2)\frac{\Pi_{i=1}^p\Gamma(a_i+z_1+z_2)}{\Pi_{j=1}^q\Gamma(c_j+z_1+z_2)}\frac{\Gamma(b+b'+z_1+z_2)\Gamma(b+z_1)}{\Gamma(b+b'+z_1)}\label{KdFtypeMB}
\end{equation}
which yields Srivastava's generalization of Carlson's identity (see \cite{Srivastava} and references therein)
\begin{equation}
	\sum_{m,n=0}^\infty\frac{
	x^{m}y^{n}}{m!n!} 
	c_{m+n}(b)_{m}(b')_n=\sum_{m,n=0}^\infty\frac{
	(x-y)^{m}y^{n}}{m!n!} 
	c_{m+n}\frac{(b'+b)_{m+n}(b)_m}{(b'+b)_m}\label{Sri_Carlson}
\end{equation}
where $c_{m+n}$ are appropriate ratios of Gamma functions, with argument depending on $m+n$.

Instead of Eq.(\ref{2F1MB1}) and the inverse change of variables applied to derive Eq.(\ref{CarlsonMB}) one can use Eq.(\ref{2F1MB3}) and the change of variables $s_1=z_1, s_2=-a-z_2$. This yields the alternative MB integral
\begin{multline}
	F_1(a,b,b';c;x,y)=\frac{\Gamma(c)}{\Gamma(a)\Gamma(b')}\left(\frac{x}{y}\right)^{b'}\int_{-i\infty}^{i\infty}\frac{dz_1}{2i\pi}\int_{-i\infty}^{i\infty}\frac{dz_2}{2i\pi}
	\left(\frac{x-y}{y}\right)^{z_1}(-x)^{z_2} \\
	\times \Gamma(-z_1)\Gamma(-z_2) 
	\frac{\Gamma(a+z_2)\Gamma(b'+b+z_1+z_2)\Gamma(b'+z_1)}{\Gamma(c+z_2)\Gamma(b'+b+z_1)}\label{F1F2MB}
\end{multline}
where we recognize the MB representation of the Appell $F_2$ function \cite{KdF} leading to the well-known transformation (see Eq.(9) p.36 in \cite{KdF} or Eq.(105) in \cite{Srivastava})
\begin{equation}
	F_1(a,b,b';c;x,y)=\left(\frac{x}{y}\right)^{b'}F_2\left(b'+b,b',a,b'+b,c;1-\frac{x}{y},x \right)\label{F1F2KdF}
\end{equation}
Once again, following the same calculations on the generalized MB integral of Eq.(\ref{F1typeMB}) gives
\begin{multline}
	I_{p,q}=\Gamma(b)(-y)^{-b'}(-x)^{b'}\int_{-i\infty}^{i\infty}\frac{dz_1}{2i\pi}
	\int_{-i\infty}^{i\infty}\frac{dz_2}{2i\pi}\left(\frac{x-y}{y}\right)^{z_1}(-x)^{z_2}	
	\Gamma(-z_1)\Gamma(-z_2)\\
	\times\frac{\Pi_{i=1}^p\Gamma(a_i+z_2)}{\Pi_{j=1}^q\Gamma(c_j+z_2)}\frac{\Gamma(b+b'+z_1+z_2)\Gamma(b'+z_1)}{\Gamma(b+b'+z_1)}\label{F2typeMB}
\end{multline}
called "$F_2$-type" and from which one can obtain a generalization of Eq.(\ref{F1F2KdF}) given as
\begin{equation}
	\sum_{m,n=0}\infty\frac{x^{m}y^{n}}{m!n!} c_{m+n}(b)_{m}(b')_n = \left(\frac{x}{y}\right)^{b'}\sum_{m,n=0}^\infty\frac{
	\left(1-\frac{x}{y}\right)^{m}x^{n}}{m!n!} c_{n}\frac{(b'+b)_{m+n}(b')_m}{(b'+b)_m}
\end{equation}

It is clear from the results shown in this section that, even for double hypergeometric functions, the number of transformed MB representations can be important. As this number can highly increase with the number of variables of the studied hypergeometric function (or multifold MB integral), this justifies the definition of a notation in order to ease the manipulation of these MB integrals, as well as to describe the calculational steps that are performed to derive them. We discuss such a notation in the next section. 

\subsection{A notation for the transformed MB representations}

Let us now explain the notation allowing us to classify the different MB integrals that can be obtained after application of the linear transformations considered in this paper.\begin{enumerate}
	\item The notation of a transformed MB integral is a concatenation of symbols which starts by the usual notation of the object or hypergeometric function that it represents (example: for the Appell $F_1$ function the usual notation is $F_1$) or, more generally, by the name given to the starting point MB integral.
	\item  This first step is followed by the adjunction of an hyphen and the label $i$ which corresponds to the $dz_i$ integral on which the transformation is applied. In order to avoid the inclusion of changes of variables in the notation when using Eqs. (\ref{CarlsonMB}), (\ref{F1typeMB}), (\ref{F1F2MB}), (\ref{KdFtypeMB}) and (\ref{F2typeMB}), one can instead write the labels of the two impacted integrals, the order of the labels being important to know how the chosen transformation acts on each variable. 
	\item One now needs a symbol to code each transformation.
For this we associate a letter to each of the five MB integrals of Section \ref{lin2F1}, as well as to Eq.(\ref{F1typeMB}), Eq.(\ref{KdFtypeMB}) and Eq.(\ref{F2typeMB}) (these are the only transformations considered in this paper). The code of a transformation is composed of the letter in lowercase associated with the integral representation of the ${}_2F_1$, $F_1$-type, $KdF$-type or $F_2$-type as it stands in the starting point MB representation, concatenated with the capital letter of the integral representation of the chosen transformation. The letters are a(A) for Eq.(\ref{2F1MB1}), b(B) for Eq.(\ref{2F1MB3}), c(C) for Eq.(\ref{2F1MB2}), d(D) for Eq.(\ref{2F1MB4}) and e(E) for Eq.(\ref{2F1MB5}). The Carlson's type transformations Eqs.(\ref{F1typeMB}), (\ref{KdFtypeMB}) and (\ref{F2typeMB}) are associated to k(K), l(L) and m(M).
	\item When multiple transformations are applied, the notation contains, from left to right, all the codes of the transformations in the order they were applied, with the insertion of the integrals labels on which the transformations are performed.
\end{enumerate}

Here are some examples to ease the understanding of this notation: the MB integral of Eq.(\ref{F1MB2}) is $F_1-2aE$, the one of Eq.(\ref{F1MB3}) is $F_1-2aE1aE$. The r.h.s of Eq.(\ref{F1MB4}) is $F_1-2aC1aE$.

Even though the notation gives a unique path, it may happen that a transformed MB integral can be obtained following different paths (for instance $F_1-1aE2aE$ is equivalent to $F_1-2aE1aE$), and thus can have different names. As a result, the aim of this notation is to give the reader a path to prove a given transformation.
	
As a last remark, we emphasize that, depending on the order of the Gamma functions in the MB integral, one may face issues with the notation. To solve these, the Gamma functions of the integrand must be written in a specific order, described as follows : 
\begin{enumerate}[label=\roman*)]
	\item The $\Gamma(-z_i)$, if any, are put to the left of any other $\Gamma$ function of the integrand (and following the increasing order of $i$ if there are several $\Gamma$ functions of this type).	
	\item $\Gamma$ functions with a lower number of $z_i$ in their argument are put to the left of $\Gamma$ functions having more $z_i$.	
	\item If one faces an integral with two or more $\Gamma$ functions having the same number of $z_i$, then : \\
		- The $\Gamma$ functions with smallest $i$ must be written first.\\
		- For a given $z_i$, the $\Gamma$ functions with a negative (smaller) coefficient in front of $z_i$ are put to the left of $\Gamma$ functions with a positive (bigger) coefficient.	\\
	\item $\Gamma$ functions without constants or parameters are put to the left of $\Gamma$ functions with constants or parameters if they have the same variables (if several of these $\Gamma$ functions have parameters, then the alphabetic order is used).
\end{enumerate}
	We give here an example in order to illustrate some of the various possibilities : \\
	$\Gamma(-z_1)\Gamma(-z_2)\Gamma(b+z_1)\Gamma(z_2)\Gamma(a+z_3)\Gamma(a+a_1+z_3)\Gamma(c+z_3)\Gamma(z_1+z_3)\Gamma(-z_2+z_3)\Gamma(z_2-z_3)\Gamma(z_2+2z_3)$

\subsection{A few other linear transformations of Appell functions}
Using the notation above, we present here a few simple and well-known results for Appell functions \cite{Srivastava, Erdelyi} that we have reproduced using their transformed MB representations.

The transformed MB integrals $F_1-2aE1aE$, $F_1-2aC1aE2cA$ (and its symmetrical $F_1-1aC2aE1cA$) and $F_1-2aC1aB2cE$ (and its symmetrical $F_1-1aC2aB1cE$) respectively produce the five Euler transformations of the Appell $F_1$ given in Eqs.(1), (2), (3) p.239 and (4), (5) p.240 of \cite{Erdelyi}:
	
	\begin{equation}
		F_1(a,b,b';c;x,y) = (1-x)^{-b}(1-y)^{-b'}F_1\left(c-a,b,b'; c; \frac{x}{x-1},\frac{y}{y-1}\right)
	\end{equation}
	\begin{equation}
		F_1(a,b,b';c;x,y) = (1-x)^{-a}F_1\left(a,c-b-b',b'; c; \frac{x}{x-1},\frac{y-x}{1-x}\right)
	\end{equation}
and its symmetrical version
\begin{equation}
		F_1(a,b,b';c;x,y) = (1-y)^{-a}F_1\left(a,b,c-b-b'; c; \frac{x-y}{1-y},\frac{y}{y-1}\right)
\end{equation}
	and
	\begin{equation}
		F_1(a,b,b';c;x,y) = (1-x)^{c-a-b}(1-y)^{-b'}F_1\left(c-a,c-b-b',b'; c; x,\frac{x-y}{1-y}\right)\label{F1MBEuler4}
	\end{equation}
and its symmetrical version
\begin{equation}
		F_1(a,b,b';c;x,y) = (1-x)^{-b}(1-y)^{c-a-b'}F_1\left(c-a,b,c-b-b'; c; \frac{y-x}{1-x},y\right)
	\end{equation}
 Similarly, the Euler transformations of the Appell $F_2$ function can be obtained from $F_2-2aD$ (and its symmetrical $F_2-1aD$) and $F_2-2aD1aD$, which respectively yield Eqs.(6), (7) and (8) p.240 of \cite{Erdelyi}:
	\begin{equation}
		F_2(a,b,b';c,c';x,y) = (1-x)^{-a}F_2\left(a,c-b,b'; c,c'; \frac{x}{x-1},\frac{y}{1-x}\right)
	\end{equation}
	\begin{equation}
		F_2(a,b,b';c,c';x,y) = (1-y)^{-a}F_2\left(a,b,c'-b'; c,c'; \frac{x}{1-y},\frac{y}{y-1}\right)
	\end{equation}
	and
	\begin{equation}
		F_2(a,b,b';c,c';x,y) = (1-x-y)^{-a}F_2\left(a,c-b,c'-b'; c,c'; \frac{x}{x+y-1},\frac{y}{x+y-1}\right)
	\end{equation}
The transformation $F_3-2aD$ produces Eq.(82) of \cite{Srivastava} 
		\begin{equation}
			F_3(a,a',b,b';c;x,y) = (1-y)^{-a'}F{}^{1:2;1}_{1:1;0}
  \left[
   \setlength{\arraycolsep}{0pt}
   \begin{array}{c@{{}:{}}c@{;{}}c}
  c-b' & a,b & a' \\[1ex]
   c & c-b' & -
   \end{array}
   \;\middle|\;
 x,\frac{y}{y-1}
 \right]
		\end{equation}
while $F_3-2aE$ yields Eq.(83) 
		\begin{equation}
			F_3(a,a',b,b';c;x,y) = (1-y)^{-b'}F{}^{1:2;1}_{1:1;0}
  \left[
   \setlength{\arraycolsep}{0pt}
   \begin{array}{c@{{}:{}}c@{;{}}c}
  c-a' & a, b & b' \\[1ex]
   c & c-a' & -
   \end{array}
   \;\middle|\;
 x,\frac{y}{y-1}
 \right]
		\end{equation}
and $F_3-2aB$ produces Eq.(84) of the same reference 
		\begin{equation}
			F_3(a,a',b,b';c;x,y) = (1-y)^{c-a'-b'}F{}^{2:2;0}_{1:2;0}
  \left[
   \setlength{\arraycolsep}{0pt}
   \begin{array}{c@{{}:{}}c@{;{}}c}
  c-a', c-b'& a, b & - \\[1ex]
   c & c-a', c-b' & -
   \end{array}
   \;\middle|\;
 x(1-y),y
 \right]
		\end{equation}
 The transformation $F_4-1aD$ produces Eq.(93) of \cite{Srivastava} :
\begin{equation}
	F_4(a,b;c,c';x,y) = (1-x)^{-a}\sum_{m,n}\frac{(x/(x-1))^m (y/(x-1))^n}{m!n!} 
	\frac{(a)_{m+n}(c-b)_{m-n}(b)_n(b-c+1)_n}{(c)_m(c')_n}
\end{equation}
while $F_4-1aE$ yields Eq.(94) 
\begin{equation}
	F_4(a,b;c,c';x,y) = (1-x)^{-b}\sum_{m,n}\frac{(x/(x-1))^m (y/(x-1))^n}{m!n!} 
	\frac{(b)_{m+n}(c-a)_{m-n}(a)_n(a-c+1)_n}{(c)_m(c')_n}
\end{equation}	
and $F_4-1aB$ Eq.(95) of \cite{Srivastava} :
\begin{equation}
	F_4(a,b;c,c';x,y) = (1-x)^{c-a-b}\sum_{m,n}\frac{x^m \left(y/(1-x)^2\right)^n}{m!n!} 
	\frac{(c-a)_{m-n}(c-b)_{m-n}(a)_n(a-c+1)_n(b)_n(b-c+1)_n}{(c)_m(c')_n}
\end{equation}

 As a last example, $F_1-2aC1aB2cA$ yields the following transformation
	
	 \begin{equation}
			F_1(a,b,c;d;x,y) = (1-x)^{d-c-b-a}F_3\left(d-a,a;d-c-b,c;d;x,\frac{x-y}{x-1}\right)
		\end{equation}
		which is a combination of Eqs.(\ref{F1MBEuler4}) and (\ref{F1F3}).

\section{Linear transformations of Srivastava's $H_C$ triple hypergeometric function \label{linHC}}
\begin{figure}
\begin{scriptsize}
\begin{center}
\begin{tikzpicture}
\begin{feynman}[large]
\vertex (a) {\(H_C\)};
\vertex [above right=of a] (aintkm);
\vertex [above left=of aintkm] (ak) {\(H_C-23lK\)};
\vertex [right=of aintkm] (am) {\(H_C-23lM\)};
\vertex [below right=of a] (c) {\(H_C-1aC\)};
\vertex [above right=of c] (b) {\(H_C-1aB\)};
\vertex [below left=of c] (e) {\(H_C-1aE\)};
\vertex [below right=of c] (d) {\(H_C-1aD\)};
\vertex [left=of c] (ctoleftpart){\(...\)};
\vertex [right=of d] (dinta);
\vertex [right=of dinta] (da);
\vertex [above right=of da] (dl) {\(H_C-1aD21kL\)};
\vertex [above left=of da] (dm) {\(H_C-1aD21kM\)};
\vertex [below right=of da] (dc) {\(H_C-1aD3aC\)};
\vertex [below right=of dc] (de) {\(H_C-1aD3aE\)};
\vertex [above right=of de] (deintk);
\vertex [above=of deintk] (dekl){\(H_C-1aD3aE21kL\)};
\vertex [above right=of deintk] (dekm){\(H_C-1aD3aE21kM\)};
\vertex [right=of de] (deintl);
\vertex [above right=of deintl] (delk){\(H_C-1aD3aE13lK\)};
\vertex [right=of deintl] (delm){\(H_C-1aD3aE13lM\)};
\vertex [below left=of dc] (dd) {\(H_C-1aD3aD\)};
\vertex [above right=of dc] (db) {\(H_C-1aD3aB\)};
\vertex [above=of db] (dbintlm);
\vertex [above right=of dbintlm] (dbl) {\(H_C-1aD3aB13kL\)};
\vertex [above left=of dbintlm] (dbm) {\(H_C-1aD3aB13kM\)};
\vertex [left=of dc] (dca);
\vertex [below left=of dca] (dcc) {\(H_C-1aD3aC2aC\)};
\vertex [above left=of dcc] (dcb) {\(H_C-1aD3aC2aB\)};
\vertex [below left=of dcc] (dcd) {\(H_C-1aD3aC2aD\)};
\vertex [below right=of dcc] (dce) {\(H_C-1aD3aC2aE\)};
\vertex [below=of dcd] (dcda);
\vertex [below right=of dcda] (dcdc) {\(H_C-1aD3aC2aD1aC\)};
\vertex [above right=of dcdc] (dcdb) {\(H_C-1aD3aC2aD1aB\)};
\vertex [below left=of dcdc] (dcdd) {\(H_C-1aD3aC2aD1aD\qquad\)};
\vertex [below right=of dcdc] (dcde) {\(H_C-1aD3aC2aD1aE\)};
\vertex [below right=of dce] (dceintc);
\vertex [right=of dceintc] (dceintintc);
\vertex [right=of dceintintc] (dcec);
\vertex [above left=of dcec] (dcea) {\(H_C-1aD3aC2aE3cA\)};
\vertex [above right=of dcec] (dceb) {\(H_C-1aD3aC2aE3cB\)};
\vertex [below left=of dcec] (dced) {\(H_C-1aD3aC2aE3cD\)};
\vertex [below right=of dcec] (dcee) {\(H_C-1aD3aC2aE3cE\)};
\diagram* {
(a) -- (c) -- (b),
(a) -- (aintkm) -- (ak),
(aintkm) -- (am),
(c) -- (d),
(c) -- (e),
(c) -- [very thick](ctoleftpart),
(d) -- (dinta),
(dinta) -- (da),
(da) -- (dl),
(da) -- (dm),
(da) -- (dc),
(dc) -- (db),
(db) -- (dbintlm),
(dbintlm) -- (dbl),
(dbintlm) -- (dbm),
(dc) -- (de),
(de) -- (deintk),
(deintk) -- (dekl),
(deintk) -- (dekm),
(de) -- (deintl),
(deintl) -- (delm),
(deintl) -- (delk),
(dc) -- (dd),
(dc) -- (dca),
(dca) -- (dcc),
(dcc) -- (dcd),
(dcc) -- (dcb),
(dcc) -- (dce),
(dcd) -- [scalar] (dcda),
(dcda) -- [scalar] (dcdc),
(dcdc) -- [scalar] (dcdb),
(dcdc) -- [scalar] (dcdd),
(dcdc) -- [scalar] (dcde),
(dce) -- [scalar] (dceintc),
(dceintc) -- [scalar] (dceintintc),
(dceintintc) -- [scalar] (dcec),
(dcec) -- [scalar] (dcea),
(dcec) -- [scalar] (dceb),
(dcec) -- [scalar] (dced),
(dcec) -- [scalar, small] (dcee),
};
\end{feynman}
\end{tikzpicture}
\end{center}
\end{scriptsize}
\caption{Half-Map 1 of $H_C$ linear transformations (the thick solid half-line with three dots is the link to Half-Map 2). Dashed lines indicate transformations that have not been computed explicitly in this paper.\label{Map1}}
\end{figure}
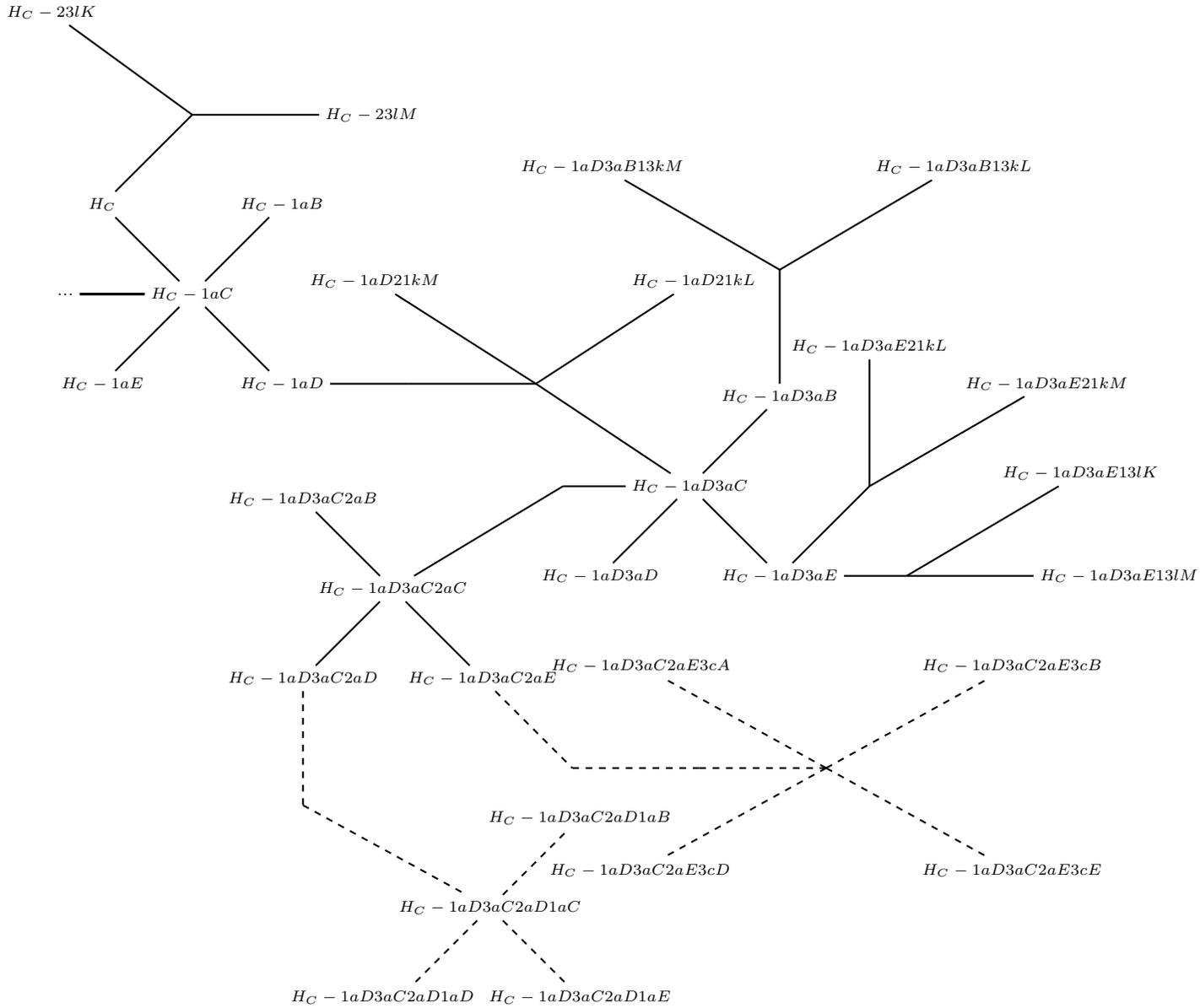

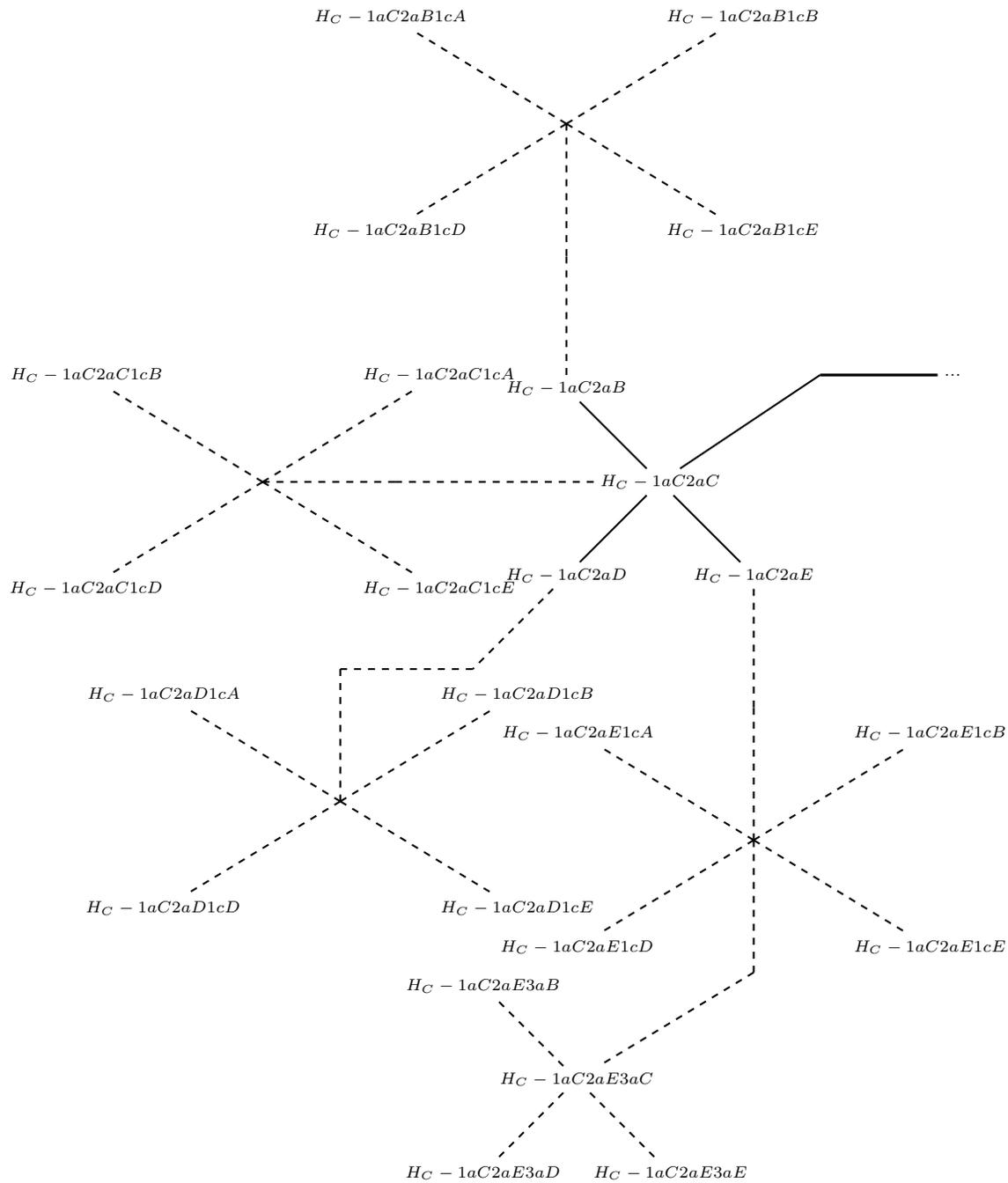
\begin{figure}
\begin{scriptsize}
\begin{center}
\begin{tikzpicture}
\begin{feynman}[large]
\vertex (c) {\( ...\)};
\vertex [left=of c] (ca);
\vertex [below left=of ca] (cc) {\(H_C-1aC2aC\)};
\vertex [above left=of cc] (cb) {\(H_C-1aC2aB\)};
\vertex [below left=of cc] (cd) {\(H_C-1aC2aD\)};
\vertex [below right=of cc] (ce) {\(H_C-1aC2aE\)};
\vertex [above=of cb] (cbintc) ;
\vertex [above=of cbintc] (cbc) ;
\vertex [above left=of cbc] (cba) {\(H_C-1aC2aB1cA\)};
\vertex [above right=of cbc] (cbb) {\(H_C-1aC2aB1cB\)};
\vertex [below left=of cbc] (cbd) {\(H_C-1aC2aB1cD\)};
\vertex [below right=of cbc] (cbe) {\(H_C-1aC2aB1cE\)};
\vertex [left=of cc] (ccintc);
\vertex [left=of ccintc] (ccintintc);
\vertex [left=of ccintintc] (ccc);
\vertex [above left=of ccc] (ccb) {\(H_C-1aC2aC1cB\)};
\vertex [above right=of ccc] (cca) {\(H_C-1aC2aC1cA\)};
\vertex [below left=of ccc] (ccd) {\(H_C-1aC2aC1cD\)};
\vertex [below right=of ccc] (cce) {\(H_C-1aC2aC1cE\)};
\vertex [below left=of cd] (cdintc);
\vertex [left=of cdintc] (cdintintc);
\vertex [below=of cdintintc] (cdc);
\vertex [above left=of cdc] (cda){\(H_C-1aC2aD1cA\)};
\vertex [above right=of cdc] (cdb){\(H_C-1aC2aD1cB\)};
\vertex [below left=of cdc] (cdd){\(H_C-1aC2aD1cD\)};
\vertex [below right=of cdc] (cde){\(H_C-1aC2aD1cE\)};
\vertex [below=of ce] (ceinta);
\vertex [below=of ceinta] (ceintinta);
\vertex [above left=of ceintinta] (cea') {\(H_C-1aC2aE1cA\)};
\vertex [above right=of ceintinta] (ceb') {\(H_C-1aC2aE1cB\)};
\vertex [below left=of ceintinta] (ced') {\(H_C-1aC2aE1cD\)};
\vertex [below right=of ceintinta] (cee') {\(H_C-1aC2aE1cE\)};
\vertex [below=of ceintinta] (cea);
\vertex [below left=of cea] (cec) {\(H_C-1aC2aE3aC\)};
\vertex [above left=of cec] (ceb) {\(H_C-1aC2aE3aB\)};
\vertex [below left=of cec] (ced) {\(H_C-1aC2aE3aD\)};
\vertex [below right=of cec] (cee) {\(H_C-1aC2aE3aE\)};
\diagram* {
(c) -- [very thick](ca),
(ca) -- (cc),
(cc) -- (cb),
(cb) -- [scalar] (cbintc),
(cbintc) -- [scalar] (cbc),
(cbc) -- [scalar](cba),
(cbc) -- [scalar](cbb),
(cbc) -- [scalar](cbd),
(cbc) -- [scalar](cbe),
(cc) -- (cd),
(cc) -- (ce),
(cc) -- [scalar] (ccintc),
(ccintc) -- [scalar] (ccintintc),
(ccintintc) -- [scalar] (ccc),
(ccc) -- [scalar] (cca),
(ccc) -- [scalar] (ccb),
(ccc) -- [scalar] (ccd),
(ccc) -- [scalar] (cce),
(cd) -- [scalar] (cdintc),
(cdintc) -- [scalar] (cdintintc),
(cdintintc) -- [scalar] (cdc),
(cdc) -- [scalar] (cda),
(cdc) -- [scalar] (cdb),
(cdc) -- [scalar] (cdd),
(cdc) -- [scalar] (cde),
(ce) -- [scalar] (ceinta),
(ceinta) -- [scalar] (ceintinta),
(ceintinta) -- [scalar] (cea'),
(ceintinta) -- [scalar] (ceb'),
(ceintinta) -- [scalar] (ced'),
(ceintinta) -- [scalar] (cee'),
(ceintinta) -- [scalar] (cea),
(cea) -- [scalar] (cec),
(cec) -- [scalar] (ceb),
(cec) -- [scalar] (ced),
(cec) -- [scalar] (cee)
};
\end{feynman}
\end{tikzpicture}
\end{center}
\end{scriptsize}
\caption{Half-Map 2 of $H_C$ linear transformations (the thick solid half-line with three dots is the link to Half-Map 1). Dashed lines indicate transformations that have not been computed explicitly in this paper.\label{Map2}}
\end{figure}

A large set of linear transformations of Srivastava's $H_C$ triple hypergeometric function can be obtained using the procedure that we have described in the preceding section. As shown in Fig. \ref{Map1} and \ref{Map2} the total number of transformed MB representations of $H_C$ is bigger than fifty, each leading to more than 10 linear transformations in general. Therefore, we only list below a few of the latter, selecting those that contain at least one triple hypergeometric series of order 2 (since these series have been studied in \cite{Srivastava}). Other results can be found in the ancillary \textit{Mathematica} file provided with this paper in the arXiv submission. The computation of the 3-fold MB integrals of this section rests on the technique presented in \cite{Ananthanarayan:2020fhl} and its related  \textsc{MBConicHulls} \textit{Mathematica} package.


Our starting-point is the MB integral representation of $H_C$, given as
\begin{displaymath}
	H_C(a,b,c;d;x,y,z)=\frac{\Gamma(d)}{\Gamma(a)\Gamma(b)\Gamma(c)}\int_{-i\infty}^{i\infty}\frac{dz_1}{2i\pi}
	\int_{-i\infty}^{i\infty}\frac{dz_2}{2i\pi}\int_{-i\infty}^{i\infty}\frac{dz_3}{2i\pi}(-x)^{z_1}(-y)^{z_2}(-z)^{z_3}	
\end{displaymath}
\begin{equation}	
	\times \Gamma(-z_1)\Gamma(-z_2)\Gamma(-z_3)\frac{\Gamma(a+z_1+z_2)\Gamma(b+z_2+z_3)\Gamma(c+z_3+z_1)}{\Gamma(d+z_1+z_2+z_3)}
\end{equation}
	This MB representation gives many building-blocks yielding series of order 2 but since its evaluation has already been performed in \cite{ABFG2}, we do not list them here. \subsection{$H_C-1aC$} 
\begin{displaymath}
	H_C(a,b,c;d;x,y,z)=\frac{\Gamma(d)}{\Gamma(a)\Gamma(b)\Gamma(c)}\int_{-i\infty}^{i\infty}\frac{dz_1}{2i\pi}
	\int_{-i\infty}^{i\infty}\frac{dz_2}{2i\pi}\int_{-i\infty}^{i\infty}\frac{dz_3}{2i\pi}(1-x)^{z_1}(-y)^{z_2}(-z)^{z_3}
\end{displaymath}
\begin{equation}	
	\times \Gamma(-z_1)\Gamma(-z_2)\Gamma(-z_3)\Gamma(a+z_1+z_2)\Gamma(b+z_2+z_3)\Gamma(c+z_3+z_1)\frac{\Gamma(d-a-c-z_1)}{\Gamma(d-a+z_3)\Gamma(d-c+z_2)}
\end{equation}
This transformed MB gives, among others, two building blocks that produce series of order 2 : $B_{1,2,3}$ and $B_{2,3,7}$. They yield the particularly simple linear transformation
\begin{multline}
	H_C(a,b,c;d;x,y,z)=\frac{\Gamma(d)\Gamma(d-a-c)}{\Gamma(d-a)\Gamma(d-c)}H_B\left(a,b,c;1-a-c+d,d-a,d-c;1-x,y,z\right) \\ +(1-x)^{d-a-c}\frac{\Gamma(d)\Gamma(a+c-d)}{\Gamma(a)\Gamma(c)}H_B\left(b,d-a,d-c;d-c,d-a,1-a-c+d;y,z,1-x\right)
\end{multline}
where $H_B$ is another triple hypergeometric function defined by Srivastava (see \cite{Srivastava} and references therein).
\subsection{$H_C-1aD$} 
\begin{displaymath}
	H_C(a,b,c;d;x,y,z)=\frac{\Gamma(d)}{\Gamma(a)\Gamma(b)\Gamma(c)}(1-x)^{-a}
	\int_{-i\infty}^{i\infty}\frac{dz_1}{2i\pi}\int_{-i\infty}^{i\infty}\frac{dz_2}{2i\pi}\int_{-i\infty}^{i\infty}\frac{dz_3}{2i\pi} \left(\frac{-x}{x-1}\right)^{z_1}\left(\frac{-y}{1-x}\right)^{z_2}(-z)^{z_3}
\end{displaymath}
\begin{equation}	
	\times \Gamma(-z_1)\Gamma(-z_2)\Gamma(-z_3)
	\frac{\Gamma(a+z_1+z_2)\Gamma(b+z_2+z_3)}{\Gamma(d+z_1+z_2+z_3)}\frac{\Gamma(c+z_3)\Gamma(d-c+z_2+z_1)}{\Gamma(d-c+z_2)}
\end{equation}
This transformation gives three building-blocks that produce series of order 2 : $B_{2,3,4}$, $B_{2,3,7}$, $B_{2,4,6}$. \begin{enumerate}
	\item From $B_{2,3,4}$ and $B_{2,3,7}$ one gets
	\begin{multline}
	H_C(a,b,c;d;x,y,z)=\frac{\Gamma(d-a-c)\Gamma(d)}{\Gamma(d-a)\Gamma(d-c)}(-x)^{-a}S_{10c}\left(a,b,c,1-d+a;1-d+a+c,d-c;\frac{x-1}{x},\frac{y}{x},z\right) \\
	+ \frac{\Gamma(a+c-d)\Gamma(d)}{\Gamma(a)\Gamma(c)}(1-x)^{d-a-c}(-x)^{c-d}S_{10c}\left(d-c,b,c,1-c;1-a-c+d,d-c;\frac{x-1}{x},\frac{y}{x} ,z\right) 
	\end{multline}
	\item From $B_{2,4,5}$, $B_{2,4,6}$ and $B_{2,5,7}$ one finds
	\begin{multline}
	H_C(a,b,c;d;x,y,z)=(1-x)^{-a} \times \\
	 \left[	\frac{\Gamma(d-a-c)\Gamma(d)\Gamma(c-b)}{\Gamma(c)\Gamma(d-c)\Gamma(d-a-b)}\left(\frac{-x}{x-1}\right)^{-a}(-z)^{-b}\sum_{m,n,p}\frac{\left(\frac{-y}{zx}\right)^m \left(\frac{x-1}{x}\right)^{n}\left(\frac{1}{z}\right)^{p}}{m!n!p!}\frac{(1-d+a+b)_{m+n+p} (a)_{m+n} (b)_{m+p}}{(1-c+b)_{n+p}(d-c)_m (1-d+a+c)_n } \right. \\
	+ \frac{\Gamma(d-a-c)\Gamma(b-c)\Gamma(d)}{\Gamma(d-c)\Gamma(d-a-c)\Gamma(b)}\left(\frac{-x}{x-1}\right)^{-a}(-z)^{-c}S_{10c}\left( a,1-d+a+c,c,b-c ; d-c,1-d+a+c ;\frac{-y}{x},\frac{x-1}{x} ,\frac{1}{z}\right)\\
	+\left. \frac{\Gamma(d)\Gamma(a+c-d)}{\Gamma(a)\Gamma(c)}\left(\frac{-x}{x-1}\right)^{c-d}(-z)^{-b}\sum_{m,n,p}\frac{\left(-\frac{y}{zx}\right)^m\left(\frac{1}{z}\right)^{n}\left(\frac{x-1}{x}\right)^{p}}{m!n!p!}\frac{(1-c+b)_{m+n+p}(b)_{m+n}(d-c)_{m+p}}{(1-c+b)_{m+n}(d-c)_m (1-a-c+d)_p} \right ]
	\end{multline}
\end{enumerate}
where $S_{10c}$ is a triple hypergeometric series defined in \cite{Srivastava}.
\subsection{$H_C-1aD3aC$}
\begin{multline}
	H_C(a,b,c;d;x,y,z)=\frac{\Gamma(d)}{\Gamma(a)\Gamma(b)\Gamma(c)}(1-x)^{-a}
	\int_{-i\infty}^{i\infty}\frac{dz_1}{2i\pi}\int_{-i\infty}^{i\infty}\frac{dz_2}{2i\pi}\int_{-i\infty}^{i\infty}\frac{dz_3}{2i\pi}
	 \left(\frac{-x}{x-1}\right)^{z_1}\left(\frac{-y}{1-x}\right)^{z_2}(1-z)^{z_3} \\
	\times \Gamma(-z_1)\Gamma(-z_2)\Gamma(-z_3)\Gamma(a+z_1+z_2)\Gamma(b+z_2+z_3)
	\frac{\Gamma(c+z_3)\Gamma(d-c-b+z_1-z_3)}{\Gamma(d-b+z_1)\Gamma(d-c+z_2)}
\end{multline} 
This transformation gives two building-blocks that produce series of order 2 : $B_{1,2,3}$ and $B_{1,2,6}$.
\begin{enumerate}
	\item  From $B_{1,2,3}$ and $B_{1,2,7}$ one obtains
	\begin{multline}
	H_C(a,b,c;d;x,y,z)=(1-x)^{-a} \\
	\times \left [ \frac{\Gamma(d)\Gamma(d-b-c)}{\Gamma(d-b)\Gamma(d-c)} S_{10c}\left(a,b,c,d-b-c;d-b,d-c;\frac{x}{x-1},\frac{y}{1-x},1-z\right) \right. \\
	+ \left. (1-z)^{c+b-d}\frac{\Gamma(d)\Gamma(d-b-c)}{\Gamma(b)\Gamma(c)} \sum_{m,n,p}\frac{\left(\frac{x(1-z)}{x-1}\right)^m\left(\frac{y}{(1-x)(z-1)}\right)^n(1-z)^{p}}{m!n!p!}\frac{(a)_{m+n}(d-b)_{m+p}(d-c)_{m+n+p}}{(1-b-c+d)_{m+p}(d-c)_{n}(d-b)_m} \right ]
	\end{multline}
	\item  From $B_{1,2,5}$ and $C_{1,2,6}$ one gets
	\begin{multline}
	H_C(a,b,c;d;x,y,z)=(1-x)^{-a}\\
	 \times \left[	(1-z)^{-b}\frac{\Gamma(d)\Gamma(c-b)}{\Gamma(c)\Gamma(d-b)} \sum_{m,n,p}\frac{\left(\frac{x}{x-1}\right)^m\left(\frac{y}{(1-x)(1-z)}\right)^n \left(\frac{1}{1-z}\right)^{p}}{m!n!p!}\frac{(a)_{m+n}(b)_{n+p}(d-c)_{m+n+p}}{(1-c+b)_{n+p}(d-c)_{n}(d-b)_{m}} \right. \\
	+\left. (1-z)^{-c}\frac{\Gamma(d)\Gamma(b-c)}{\Gamma(b)\Gamma(d-c)} S_{10c}\left(a,d-b,c,b-c;d-c,d-b;\frac{y}{(1-x)},\frac{x(1-z)}{x-1},\frac{1}{1-z}\right) \right ]
	\end{multline}
\end{enumerate}
\subsection{$H_C-1aD3aE$} 
\begin{multline}
	H_C(a,b,c;d;x,y,z)=\frac{\Gamma(d)}{\Gamma(a)\Gamma(b)}(1-x)^{-a}(1-z)^{-b} \\
	\times \int_{-i\infty}^{i\infty}\frac{dz_1}{2i\pi}\int_{-i\infty}^{i\infty}\frac{dz_2}{2i\pi} \int_{-i\infty}^{i\infty}\frac{dz_3}{2i\pi}
	\left(\frac{-x}{x-1}\right)^{z_1}\left(\frac{-y}{(1-x)(1-z)}\right)^{z_2}\left(\frac{-z}{z-1}\right)^{z_3}		
	\\ \times \Gamma(-z_1)\Gamma(-z_2)\Gamma(-z_3)
	\frac{\Gamma(a+z_1+z_2)\Gamma(b+z_2+z_3)}{\Gamma(d+z_1+z_2+z_3)}\frac{\Gamma(d-c+z_1+z_2+z_3)}{\Gamma(d-c+z_2)}
\end{multline}
This transformation gives eight building blocks that produce series of order 2 : $B_{1,2,5}$, $B_{1,2,6}$, $B_{1,4,6}$, $B_{2,3,4}$, $B_{2,3,6}$, $B_{2,4,6}$, $B_{2,5,6}$ and $B_{3,5,6}$. In order to ease the reading of these results, we provide here a list of the building blocks obtained during the calculation. \\
We start by listing building blocks of order two :
	\begin{multline}
	B_{1,2,5}=(1-x)^{-a}z^{-b}\frac{\Gamma(d-b-c)\Gamma(d)}{\Gamma(d-b)\Gamma(d-c)}S_{10h}\left(a,b,d-b-c,1-d+b;d-c;\frac{x}{1-x},\frac{y}{z(x-1)},\frac{1-z}{z}\right)
	\end{multline}
	where $S_{10h}$ is a series defined in \cite{Srivastava} and associated to the index $10h$.
	\begin{multline}
	B_{1,2,6}=(1-x)^{-a} z^{c-d}(1-z)^{d-b-c}\frac{\Gamma(b+c-d)\Gamma(d)}{\Gamma(b)\Gamma(c)} \\
	\times F_{14}\left( d-c,a,1-c;d-c,1-b-c+d;\frac{y}{z(x-1)},-\frac{x(z-1)}{z(x-1)},\frac{z-1}{z}\right)
	\end{multline}
	where $F_{14}$ is one of Lauricella's functions associated to the index $21a$ in \cite{Srivastava}.
	\begin{multline}
	B_{2,3,4}=(1-z)^{-b} x^{-a}\frac{\Gamma(d-a-c)\Gamma(d)}{\Gamma(d-a)\Gamma(d-c)} \\
	\times S_{10h}\left(b,a,d-a-c,1-d+a;d-c ; \frac{z}{1-z},\frac{y}{x(z-1)},\frac{1-x}{x}\right)
	\end{multline}
	One will note that $B_{2,3,4}$ is the symmetric of $B_{1,2,5}$.
	\begin{multline}
	B_{2,3,6}=(1-z)^{-b}x^{c-d}(1-x)^{d-a-c}\frac{\Gamma(a+c-d)\Gamma(d)}{\Gamma(a)\Gamma(c)} \\ 
	\times F_{14}\left(d-c,b,1-c;d-c,1-a-c+d;\frac{y}{x(z-1)},\frac{-z(x-1)}{x(z-1)},\frac{x-1}{x}\right)
	\end{multline}
	$B_{2,3,6}$ is the symmetric of $B_{1,2,6}$.
	\begin{multline}
		B_{1,4,6}=\frac{\Gamma(d)\Gamma(b+c-d)}{\Gamma(b)\Gamma(c)}(1-z)^{-b}\left(\frac{-z}{y}\right)^{a}\left(\frac{-z}{z-1}\right)^{c-d} \\
		\times S_{9b}\left(a,1-d+a+c,1-c,d-a-c;1-b-c+d;\frac{z(1-x)}{y},\frac{x(1-z)}{y},\frac{z-1}{z}\right)
	\end{multline}
	where $S_{9b}$ is a series defined in \cite{Srivastava}.
		\begin{multline}
		B_{2,4,6}=\frac{\Gamma(d)\Gamma(a+b+c-d)\Gamma(d-a-c)}{\Gamma(b)\Gamma(c)\Gamma(d-c)}(1-z)^{-b}\left(\frac{x(1-z)}{z}\right)^{-a}\left(\frac{z}{1-z}\right)^{c-d} \\
		\times S_{8d}\left(a,1-c,a+b+c-d,d-a-c;d-c;\frac{z(1-x)}{x(1-z)},\frac{y}{x(1-z)},\frac{z-1}{z}\right)
	\end{multline}
	where $S_{8d}$ is a series defined in \cite{Srivastava}. 
	\begin{multline}
		B_{2,5,6}=\frac{\Gamma(d)\Gamma(a+b+c-d)\Gamma(d-b-c)}{\Gamma(a)\Gamma(c)\Gamma(d-c)}(1-x)^{-a}\left(\frac{z(1-x)}{x}\right)^{-b}\left(\frac{x}{1-x}\right)^{c-d} \\
		\times S_{8d}\left(b,1-c,a+b+c-d,d-b-c;d-c;\frac{x(1-z)}{z(1-x)},\frac{y}{z(1-x)},\frac{x-1}{x}\right)
	\end{multline}	
	$B_{2,5,6}$ is the symmetric of $B_{2,4,6}$.
	\begin{multline}
		B_{3,5,6}=\frac{\Gamma(d)\Gamma(a+c-d)}{\Gamma(a)\Gamma(c)}(1-x)^{-a}\left(\frac{-x}{y}\right)^{b}\left(\frac{-x}{x-1}\right)^{c-d} \\
		\times S_{9b}\left(b,1-d+b+c,1-c,d-b-c;1-a-c+d;\frac{x(1-z)}{y},\frac{z(1-x)}{y},\frac{x-1}{x}\right)
	\end{multline}
	$B_{3,5,6}$ is the symmetric of $B_{1,4,6}$.\\
	
Let us now list the building blocks of order 3 :
	\begin{multline}
		B_{1,3,4}=\frac{\Gamma(d)\Gamma(b-a)}{\Gamma(d-a)\Gamma(b)} (1-z)^{-b}\left(\frac{-y}{1-z}\right)^{-a} \\
		\times  \sum_{m,n,p}\frac{\left(\frac{x(1-z)}{y}\right)^m \left(\frac{z}{z-1}\right)^n\left(\frac{(1-x)(1-z)}{y}\right)^p}{m!n!p!}(a)_{m+p}(d-a-c)_{n-p}(b-a)_{n-m-p}(1-d+a+c)_{m+p}(1-d+a)_{p-n}
	\end{multline}
	\begin{multline}
		B_{1,3,5}=\frac{\Gamma(d)\Gamma(a-b)}{\Gamma(d-b)\Gamma(a)} (1-x)^{-a}\left(\frac{-y}{1-x}\right)^{-b} \\
		\times  \sum_{m,n,p}\frac{\left(\frac{z(1-x)}{y}\right)^m \left(\frac{x}{x-1}\right)^n\left(\frac{(1-x)(1-z)}{y}\right)^p}{m!n!p!}(b)_{m+p}(d-b-c)_{n-p}(a-b)_{n-m-p}(1-d+b+c)_{m+p}(1-d+b)_{p-n}
	\end{multline}
	$B_{1,3,5}$ is the symmetric of $B_{1,3,4}$.
	\begin{multline}
		B_{1,4,5}=\frac{\Gamma(d)\Gamma(d-b-c)\Gamma(b-a)}{\Gamma(b)\Gamma(d-b)\Gamma(d-a-c)} z^{-b}\left(\frac{-y}{z}\right)^{-a} \\
		\times  \sum_{m,n,p}\frac{\left(\frac{zx}{y}\right)^m \left(\frac{z-1}{z}\right)^n\left(\frac{z(1-x)}{y}\right)^p}{m!n!p!}(a)_{m+p}(1-d+a+c)_{m+p}(d-b-c)_{m-n}(b-a)_{n-m-p}(1-d+b)_{n-m}
	\end{multline}
	\begin{equation}
		B_{2,4,5}=\frac{\Gamma(d)\Gamma(d-a-b-c)}{\Gamma(d-c)\Gamma(d-a-b)} z^{-b} x^{-a} \sum_{m,n,p}\frac{\left(\frac{x-1}{x}\right)^{n}\left(\frac{y}{xz}\right)^{m}\left(\frac{z-1}{z}\right)^{p}}{m!n!p!}\frac{(1-d+a+b)_{m+n+p}(a)_{m+n}(b)_{m+p}}{(1-d+a+b+c)_{m+n+p}(d-c)_m}
	\end{equation}
	$B_{3,4,5}$ is the symmetric of $B_{1,4,5}$.
	\begin{multline}
		B_{3,4,5}=\frac{\Gamma(d)\Gamma(d-a-c)\Gamma(a-b)}{\Gamma(a)\Gamma(d-a)\Gamma(d-b-c)} x^{-a}\left(\frac{-y}{x}\right)^{-b} \\
		\times  \sum_{m,n,p}\frac{\left(\frac{zx}{y}\right)^m \left(\frac{x-1}{x}\right)^n\left(\frac{x(1-z)}{y}\right)^p}{m!n!p!}(b)_{m+p}(1-d+b+c)_{m+p}(d-a-c)_{m-n}(a-b)_{n-m-p}(1-d+a)_{n-m}
	\end{multline}
	\begin{multline}
		B_{4,5,6}=\frac{\Gamma(d)\Gamma(a+b+c-d)\Gamma(d-a-c)\Gamma(d-c-b)}{\Gamma(a)\Gamma(b)\Gamma(c)\Gamma(a-b-2c+2d)} \frac{(-y)^{d-a-b-c}}{x^{d-c-b}z^{d-c-a}} \\
		\times  \sum_{m,n,p}\frac{\left(\frac{z(1-x)}{y}\right)^m \left(\frac{x(1-z)}{y}\right)^n\left(\frac{y}{xz}\right)^p}{m!n!p!}(1-c)_p (1-a+b+2c-2d)_{m+n-p}(a+b+c-d)_{m+n-p}(d-a-c)_{p-m}(d-c-b)_{p-n}
	\end{multline}

From these building blocks, we obtain the following eight linear transformations of $H_C$ :
\begin{align}
	H_C(a,b,c;d;x,y,z)& = B_{1,2,5} + B_{1,2,6} \\
	& = B_{2,3,4} + B_{2,3,6} \\
	& = B_{1,3,4} + B_{1,3,6} + B_{3,4,5} + B_{3,5,6} \\
	& = B_{1,3,5} + B_{1,3,6} + B_{1,4,5} + B_{1,4,6} \\
	& = B_{1,2,6} + B_{2,4,5} + B_{2,5,6} \\
	& = B_{2,3,6} + B_{2,4,5} + B_{2,4,6} \\
	& = B_{1,3,6} + B_{1,4,5} + B_{1,4,6} + B_{3,4,5} + B_{3,5,6} \\
	& = B_{1,3,6} + B_{1,4,6} + B_{2,4,5} + B_{3,5,6} + B_{4,5,6}
\end{align}

\section{Conclusions}
We have shown in this paper on the example of Srivastava's $H_C$ triple hypergeometric function how, by successive applications of linear transformations of the Gauss $_2F_1$ and Appell $F_1$ functions, one can generate linearly transformed Mellin-Barnes representations for this hypergeometric function of three variables. The powerful, and so far unique, computational method of MB integrals recently developed and presented in \cite{Ananthanarayan:2020fhl} can then be efficiently used to explicitly compute large sets of linear transformations of this kind of objects. 
It is for instance straightforward to apply the same procedure to other triple hypergeometric functions such as Lauricella's $F_A, F_B, F_C, F_D$, Srivastava's $H_A$, $H_B$, etc.
We have begun the corresponding investigations of such cases. These results can be used to build \textit{Mathematica} packages dedicated to the evaluation of these functions, in the same way as this has been performed for the Appell $F_2$ case in \cite{Ananthanarayan:2021bqz}. 
Applications of this method in quantum field theory can also be potentially interesting for the evaluation of Feynman integrals.

\end{document}